\documentclass[preprints,article,accept,moreauthors,pdftex]{Definitions/mdpi} 

\usepackage{bm}
\usepackage{bbm}
\usepackage{listings}
\usepackage{subfigure}
\usepackage{amsmath}
\usepackage{amssymb}
\usepackage{amsthm}
\usepackage{booktabs}
\usepackage{hyperref}
\usepackage{erewhon} 
\usepackage[sf,scale=1]{libertine}
\usepackage{fourier}
\usepackage[T1]{fontenc}
\usepackage{algorithm}
\usepackage{algorithmic}

\hypersetup{colorlinks,citecolor=teal}

\firstpage{1} 
\pubvolume{xx}
\issuenum{1}
\articlenumber{1}
\pubyear{2020}
\copyrightyear{2020}
\history{}

\newcommand*\diff{\mathop{}\!\kern0pt\mathrm{d}}

\Title{On the Bachelier implied volatility at extreme strikes}
\Author{Fabien Le Floc'h}
\AuthorNames{Fabien Le Floc'h}
\address{}
\corres{Correspondence: fabien@2ipi.com}
\abstract{Roger Lee proved that the Black-Scholes implied variance can not grow faster than linearly in log-moneyness. This paper investigates what happens in the Bachelier (or Normal) implied volatility world, making sure to cover the various aspects of vanilla option arbitrages.}
\keyword{implied volatility, b.p. vol, Bachelier, Black-Scholes, European option, arbitrages}

\begin{document}

\section{Introduction}
\label{intro}
Under the Bachelier (also known as normal, or b.p. vol) model, a drift-less underlying asset price $F$ is subject to the following stochastic differential equation:
\begin{equation*}
	\diff F = \sigma_N \diff W\,,
\end{equation*}
where $W$ is a Brownian motion and $\sigma_N$ is the Bachelier volatility. 
The prices of European call and put options of maturity $T$ and strike $K$ are known in closed form and read \cite{bachelier1900theorie}
\begin{align}
	C_N(K,\sigma_N) &= B(0,T) \left[ (F-K)\Phi\left(\frac{F-K}{\sigma_N\sqrt{T}}\right) + \sigma_N \sqrt{T}\phi\left(\frac{F-K}{\sigma_N\sqrt{T}}\right) \right]\,,\\
	P_N(K,\sigma_N) &= B(0,T) \left[ (K-F)\Phi\left(\frac{K-F}{\sigma_N\sqrt{T}}\right) + \sigma_N \sqrt{T}\phi\left(\frac{F-K}{\sigma_N\sqrt{T}}\right) \right]\,,
\end{align}
where $\phi, \Phi$ are respectively the normal distribution and cumulative normal distribution functions and $B(0,T)$ is the discount factor to maturity.

In practice, the underlying asset is a forward rate, such as a forward swap rate for an interest rate swaption, or the spread between two commodity forward or future prices. As in the Black-Scholes model, the market prices of European options imply different values of the Bachelier volatility, depending on the strike and the time to maturity of the option considered.

Roger Lee proves that the Black-Scholes implied volatility asymptotics can not be arbitrary in \cite{lee2004moment}: the implied distribution does not have finite moments if the square of the total implied variance $w(y,T)=\sigma^2(y,T)$ grows faster than $2|y|$ where $y$ is the log-moneyness $y=\ln\frac{K}{F}$, $K$ is the option strike, $F$ the underlying asset forward to maturity $T$ and $\sigma$ is the Black-Scholes implied volatility for the given moneyness and maturity.
As a consequence, any implied variance extrapolation must be at most linear in the wings, a fact of great practical interest.

Earlier, Hodges derived simpler bounds on the Black-Scholes implied volatility based on the option slopes in \cite{hodges1996arbitrage}, later refined by Gatheral in \cite{gatheral1999volatility}. Benaim et al. \cite{benaim2009regular,benaimblack} explore the Black-Scholes implied volatility asymptotics of various models including the SABR stochastic volatility model of Hagan et al. \cite{hagan2002managing}.

With the advent of low or negative interest rates, practitioners have moved towards the use of the Bachelier model. Then, a natural question that arises is: what kind of extrapolation is acceptable in the Bachelier model?

We start with a tight upper asymptotic boundary of the Bachelier implied volatility. We continue by looking at the sufficient conditions for the existence of the distribution moments in terms of implied volatility. We conclude with a visualization of the boundary cases, both in the Black and in the Bachelier models.

\section{Asymptotic upper bound of the Bachelier implied volatility}

\begin{Theorem}
	The Bachelier implied volatility $\sigma_N$ is bounded above by $\frac{K-F}{\sqrt{2 T \ln K}}$ when $K \to +\infty$ or more precisely,
	if $\exists b < 2 \mid \forall K \in \mathbb{R}^+\,, \exists K_0 > K $ such that $\sigma_N(K_0,T) > \frac{K_0-F}{\sqrt{b T \ln K_0}}$ then the Bachelier option price $C_N(K,T,\sigma_N)$ has arbitrages.
\end{Theorem}
\begin{proof}
	In order for the model to be meaningful for option pricing, the forward price must be finite, that is $\mathbb{E}\left[F_T\right] < +\infty$. By dominated convergence, we must have 
	\begin{equation*}
	\lim\limits_{K \to\infty} C_N(K,\sigma_N(K)) = \lim\limits_{K \to\infty} \mathbb{E}[ |F_T-K|^+]= 0\,.
	\end{equation*}
	
	Let $v(K)= \frac{K-F}{\sqrt{b T\ln K}}$, we have
	\begin{equation*}
	\frac{K-F}{v(K)\sqrt{T}} = \sqrt{b \ln K}\,,\quad 	\phi\left(\frac{K-F}{v\sqrt{T}}\right) = \frac{1}{\sqrt{2\pi K^b}}
	\end{equation*}
    An expansion of the cumulative normal distribution is
    \begin{equation}\label{eqn:cnd_asymptotic}
    \Phi(-x) = \frac{\phi(x)}{x}\left[1+\sum_{n=1}^{N-1}(-1)^n \frac{(2n-1)!!}{x^{2n}}\right] + O\left(x^{-2N}\phi(x)\right) \textmd{ as } x \to \infty\,.
    \end{equation}
    When $K \to \infty$, taking the first term of the serie, we have then
	\begin{equation*}
	C_N(K,v(K)) = \frac{v^3(K)T^\frac{3}{2}}{(K-F)^2}\phi\left(\frac{K-F}{v\sqrt{T}}\right) + ...
	\end{equation*}
	or \begin{equation}\label{eqn:call_asymptotic}
	C_N(K,v(K))= \frac{K-F}{(b \ln K)^\frac{3}{2}} \sqrt{\frac{1}{2\pi K^b}} + ...
	\end{equation}
	For $b < 2$, $C_N(K,v(K)) \to \infty$.
	As $C_N(K,\sigma_N)$ is monotone in its second argument, $\sigma_N$ is bounded by $v(K)=\frac{K-F}{\sqrt{b\ln K}}$ when $K \to \infty$ for all $b<2$.
\end{proof}

The bound for negative strikes is very similar. In this case, we will rely on the put option price $P$. Under the Bachelier model, we have $P_N(K,\sigma_N) = (K-F)\Phi\left(\frac{K-F}{\sigma_N\sqrt{T}}\right) + \sigma_N\sqrt{T}\phi\left(\frac{K-F}{\sigma_N\sqrt{T}}\right)$. Let $v(K)= \frac{|K|+F}{\sqrt{bT \ln|K|}}$. When $K \to -\infty$, the expansion of the cumulative normal distribution leads to $P_N(K,v(K)) = \frac{|K|+F}{(b \ln |K|)^\frac{3}{2}} \sqrt{\frac{1}{2\pi |K|^b}} + ...$.

\section{Acceptable lower bound of the Bachelier implied volatility}
\begin{Lemma}
	If $\sigma_N(K,T) = \frac{K-F}{\sqrt{b T \ln (K-F)}}$ for $b \geq 2$, then $\exists K_0 \mid \forall K>K_0, C_N(K,T,\sigma_N(K))$ has no arbitrages.
\end{Lemma}
\begin{proof}
According to Equation \ref{eqn:call_asymptotic}, when $b\geq 2$, we have $C(K,v(K)) \to 0$ as $K \to \infty$. 

In order to preserve arbitrage, the call and put prices must also obey
 \begin{align*}
-1 &\leq \frac{\partial C}{\partial K} \leq 0\,,\\
0 &\leq \frac{\partial P}{\partial K} \leq 1\,.
\end{align*}
In terms of the variables $s=\sigma_N \sqrt{T}$ and $x=K-F$, this translates to
\begin{equation}
 -\frac{\Phi\left(\frac{x}{s}\right)}{\phi\left(\frac{x}{s}\right)} \leq \frac{\partial s}{\partial x} \leq \frac{\Phi\left(-\frac{x}{s}\right)}{\phi\left(\frac{x}{s}\right)}\,.
\end{equation}
Let us consider $r(x)=\frac{x}{\sqrt{b \ln (x)}}$, we have
\begin{align*}
 \frac{\partial r}{\partial x}(x) &= \frac{1}{\sqrt{b \ln(x)}}- \frac{b }{2 \left(b \ln (x)\right)^\frac{3}{2}}\\
 &= \frac{r}{x} - \frac{b r^3}{2 x^3}\,.
\end{align*}
According to the asymptotic expansion of the cumulative normal distribution function given by Equation \ref{eqn:cnd_asymptotic} we have for $x$ sufficiently large:
\begin{equation*}
\frac{\Phi\left(-\frac{x}{r}\right)}{\phi\left(\frac{x}{r}\right)} \geq \frac{r}{x}- \frac{r^3}{x^3} + 3 \frac{r^5}{x^5} - 15\frac{r^7}{x^7}\,.
\end{equation*}
For $b \geq 2$, we thus have 
\begin{equation*} 0 \leq \frac{\partial r}{\partial x}(x) \leq \frac{\Phi\left(-\frac{x}{s}\right)}{\phi\left(\frac{x}{s}\right)}
\end{equation*}
for $x$ sufficiently large.
\end{proof}
In similar fashion, it can be proven that a tighter lower bound is \begin{equation}\sigma_N(K,T) = \frac{K-F}{\sqrt{2T\ln(K-F)}} \left[1+ \frac{d}{\ln(K-F)}\right]\,,\end{equation} for $d \in \mathbb{R}$.


\section{Relation between moments explosion and the Bachelier implied volatility}
\begin{Theorem}
	If the Bachelier implied volatility is below $v_R(K)=\frac{K-F}{\sqrt{b T \ln K}}$ as $K \to \infty$ with $b > 2(1+p)$ and below $v_L(K)=\frac{|K|+F}{\sqrt{c T \ln |K|}}$ as $K \to -\infty$ with $c > 2(1+p)$, then the $(p+1)$-th moment $\mathbb{E}\left[F^{p+1}\right]$ exists.
\end{Theorem}
\begin{proof}
Let $f$ be a function of class $C^2$ on $\mathbb{R}$, then, following the steps of Carr and Madan in \cite[appendix A]{carr2001optimal}, we have 
\begin{equation*}
f(F) = f(S) + f'(S)(F-S) + \int_{S}^{\infty}f''(k) (F-k)^{+} dk + \int_{-\infty}^{S}f''(k) (k-F)^{+} dk\,.
\end{equation*}

Let $p \in \mathbb{N}^\star$, applying the above to $f(F)=F^{p+1}$ and $S=0$ leads to
\begin{equation*}
\mathbb{E}\left[S^{p+1}\right] = p(p+1) \left[\int_{0}^{\infty}k^{p-1}C_N(k,T,\sigma_N(k)) dk + \int_{-\infty}^{0}k^{p-1} P_N(k,T,\sigma_N(k)) dk \right]\,.
\end{equation*}
Given that the call and put prices are monotonically increasing in the volatility argument $\sigma_N$, it suffices to prove that the call integral converges for $\sigma_N(K)= v_R(K)$ and put integral converges for $\sigma_N(K)= v_L(K)$.

Now according to Equation \ref{eqn:call_asymptotic}, we have as $k \to \infty$,
\begin{align*}
k^{p-1}C_N(k,T,\sigma_N(k)) = \frac{1}{\sqrt{2\pi}}\frac{k^{p-\frac{b}{2}}}{(b \ln k)^{\frac{3}{2}}} + O\left(k^{p-\frac{b}{2}-1}\right)\,.
\end{align*}
The call integral will converge for $p-\frac{b}{2} < -1$, that is for $b > 2(1+p)$.
Similarly, the put integral will converge for $p-\frac{c}{2} < -1$, that is for $c > 2(1+p)$.
\end{proof}

\begin{Corollary}
If the Bachelier implied volatility behaves like $|K|^\beta$ as $K \to \pm \infty$, for $\beta < 1$, then all the moments exists.
\end{Corollary}

\section{Visualization of limiting cases}
\begin{figure}[H]
		\subfigure[\label{fig:bachelier_08} Bachelier]{
			\includegraphics[width=.45\textwidth]{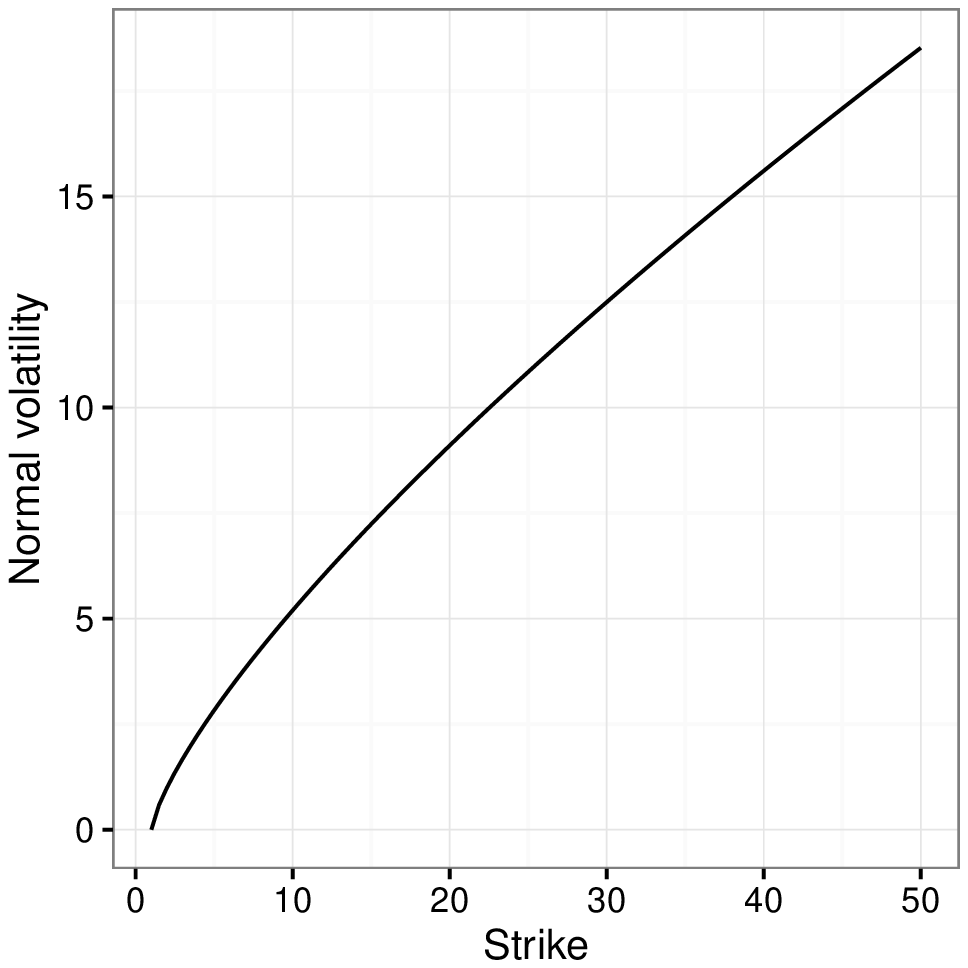}
		}
		\subfigure[\label{fig:black_08} Black]{
				\includegraphics[width=.45\textwidth]{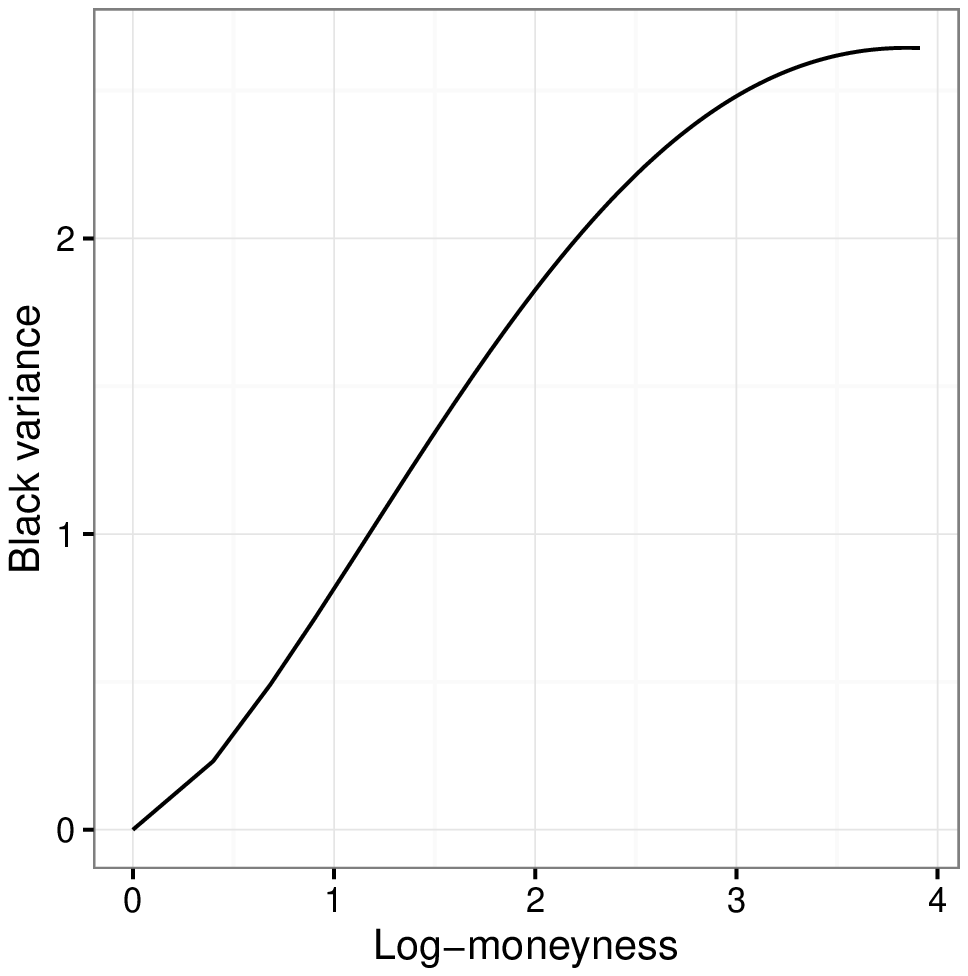}
		} 
	\caption{\label{fig:bachelier_black_08} Black variance corresponding to the Bachelier normal volatility smile $\sigma_N = (K-f)^\frac{3}{4}$.}
\end{figure}

Figure \ref{fig:bachelier_black_08} shows that the Black implied volatility will flatten in general for large strikes when the Bachelier implied volatility is of the form $\sigma_N = (K-f)^\beta$ with $\beta < 1$.

\begin{figure}[H]
		\subfigure[\label{fig:bachelier_black_1_n} Bachelier]{
			\includegraphics[width=.45\textwidth]{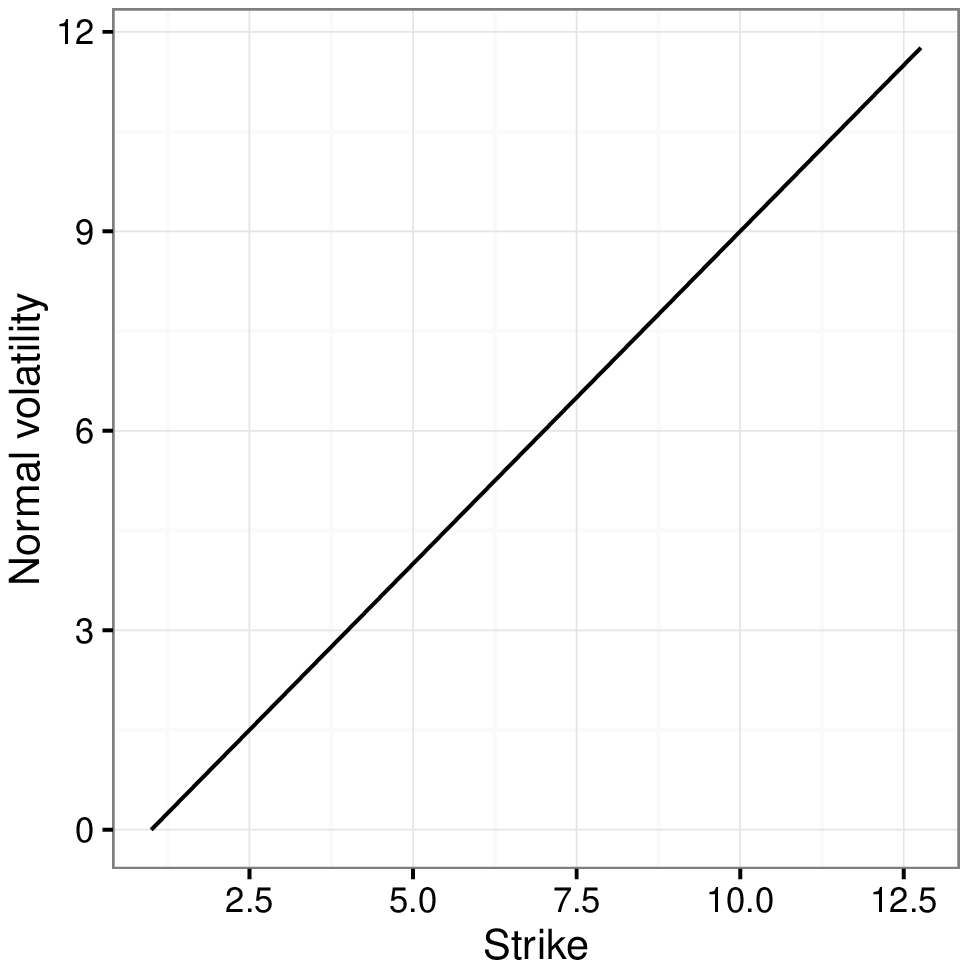}}
		\subfigure[\label{fig:bachelier_black_1_b} Black]{
			\includegraphics[width=.45\textwidth]{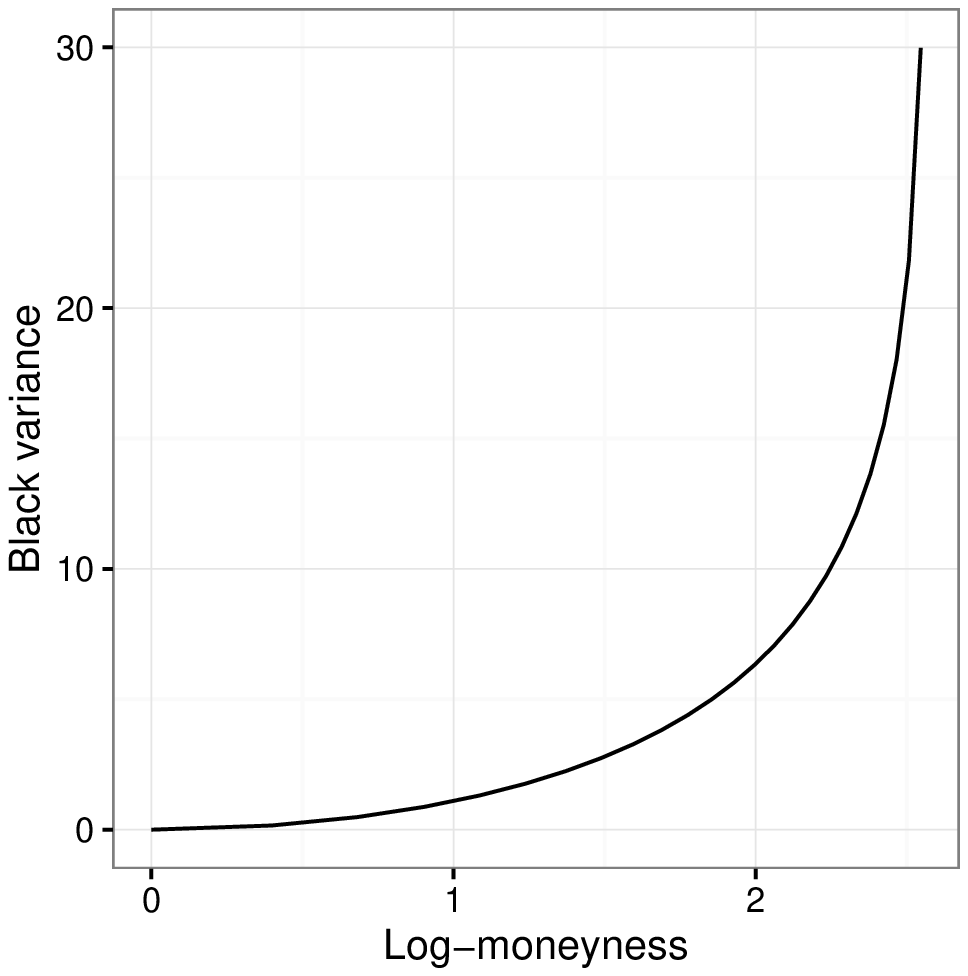}}
	\caption{\label{fig:bachelier_black_1} Black variance corresponding to the Bachelier normal volatility smile $\sigma_N = (K-f)$.}
\end{figure}

\begin{figure}[H]
		\subfigure[\label{fig:black_bachelier_05_n} Bachelier]{
			\includegraphics[width=.45\textwidth]{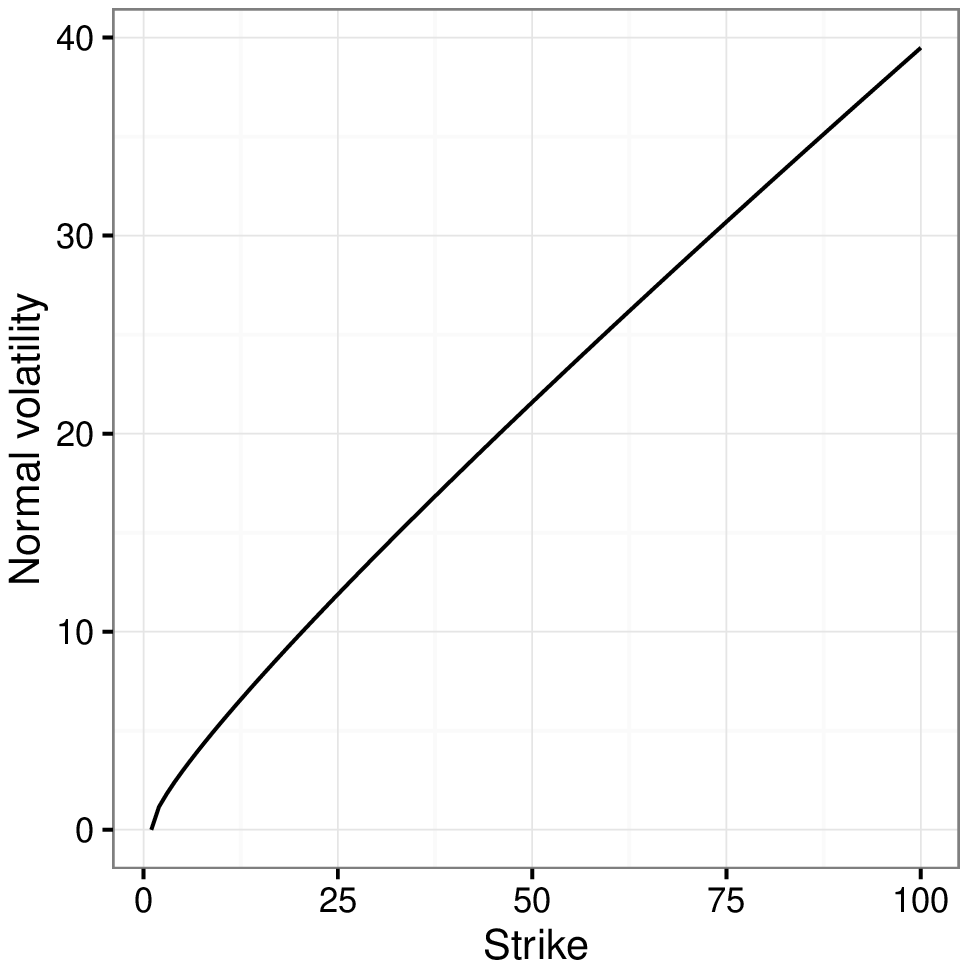}}
		\subfigure[\label{fig:black_bachelier_05_b} Black]{
			\includegraphics[width=.45\textwidth]{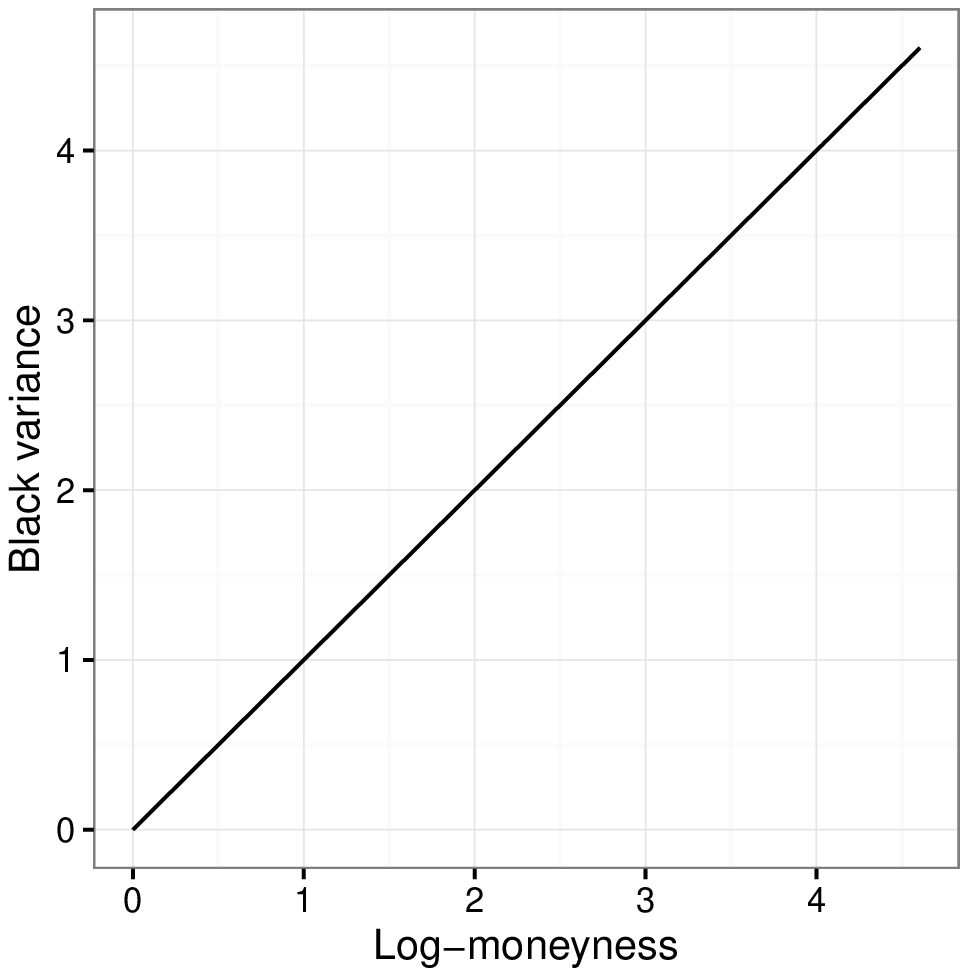}}
	\caption{\label{fig:black_bachelier_05} Bachelier normal volatility corresponding to the Black variance $w = \ln\frac{K}{F}$.}
\end{figure}

A linear normal volatility can lead to super-linear Black variance as in Figure \ref{fig:bachelier_black_1}.
The limit case for Black volatilities, a linear Black variance looks almost linear (but is not) in the wings in terms of normal volatility (Figure \ref{fig:black_bachelier_05}).

\section{Conclusion}
We have shown that the Bachelier implied volatility asymptotic for large strikes is bounded by the function $\sigma_N(K)=\frac{K-F}{\sqrt{b T \ln K}}$, above for $b < 2$ and below for $b \geq 2$ to guarantee a proper behavior of the vanilla call option prices. For $b \geq 2$, this function guarantees the absence of arbitrages. The value of $b$ determines how many moments of the distribution are finite.

We leave to further research the refinement of the boundary with additional terms. Unlike the Black-Scholes case, there is, a priori, no exact simple closed form formula for the asymptotic in the Bachelier model.

\funding{This research received no external funding.}
\conflictsofinterest{The authors declare no conflict of interest.}
\acknowledgments{Fruitful conversations with Gary Kennedy and Peter Carr.}
\externalbibliography{yes}
\bibliography{volatility_asymptotics.bib}
\appendixtitles{no}
\appendix
\section{Asymptotic expansions relating Black and Bachelier volatilities}
Although the problem of converting Black volatilities (also called basis point volatilities or Bachelier volatilities) to normal volatilities can be solved quite efficiently by relying on inverting the Black-Scholes price through the nearly closed form representation of Le Floc'h \cite{lefloch2014fast}, or if the reverse is needed, through a good Black implied volatility solver such as the one of J\"ackel \cite{jackel2013let}, the behavior around the money is well described by the following simple and fast expansions:
\begin{itemize}
	\item Hagan simple second order approximation \cite{hagan2002managing}
	\begin{equation}
	\sigma_N = \sigma_B \frac{f-K}{\ln \frac{f}{K}}\left(1+\frac{1}{24}\sigma_B^2 T + ...\right)
	\end{equation}
	\item Hagan fourth order approximation \cite{hagan2002managing} 
	\begin{equation}
	\sigma_N = \sigma_B \sqrt{fK}\frac{1+\frac{1}{24}\ln^2 \frac{f}{K}+  
		\frac{1}{1920}\ln^4 \frac{f}{K}+ ...}{1+\frac{1}{24}\left(1-\frac{1}{120}\ln^2 \frac{f}{K}\right)\sigma_B^2T+\frac{\sigma_B^4 T^2}{5760}+...}
	\end{equation}
	\item The small time $O(T^2 \ln T)$ expansion of Grunspan \cite{grunspan2011note}
	\begin{equation}
	\sigma_N = \sigma_B \frac{f-K}{\ln \frac{f}{K}}\left(1- \frac{\ln\left(\frac{1}{\sqrt{Kf}}\frac{f-K}{\ln\frac{f}{K}}\right)}{\ln^2 \frac{f}{K}} \sigma_B^2 T\right) + O(T^2 \ln T)
	\end{equation}
	Following their paper derivation, we can also find the inverse expression.
	\item The third order Black implied volatility expansion of Lorig \cite{lorig2015explicit}
	\begin{equation}
	\sigma_B = \frac{\sigma_N}{f} \left[1-\frac{1}{2}\ln \frac{K}{f}+\frac{1}{96}\left(8\ln^2 \frac{K}{f}+  \left(\frac{\sigma_N}{f}\right)^2 T(4-\left(\frac{\sigma_N}{f}\right)^2 T)\right) \right.\\\left.+ \frac{1}{192}\ln\frac{K}{f}\left(\frac{\sigma_N}{f}\right)^2 T(-12+5 \left(\frac{\sigma_N}{f}\right)^2 T) 	\right]
	\end{equation}
	
\end{itemize}
\end{document}